\documentclass[conference]{IEEEtran}
\usepackage{stmaryrd}
\usepackage{amsfonts}
\usepackage{graphics} 
\usepackage{subfigure}
\usepackage{epsfig} 
\usepackage{mathptmx} 
\usepackage{times} 
\usepackage{amssymb}  
\usepackage{floatrow}
\usepackage{multirow}
\usepackage{lettrine}
\usepackage{algorithm}
\usepackage{algpseudocode}
\usepackage{color}
\usepackage{booktabs}
\usepackage{graphicx}
\usepackage{makecell}
\usepackage{supertabular}
\usepackage{amsmath} 
\usepackage{amsmath}
\usepackage{amsfonts}
\usepackage{mathrsfs}
\usepackage{graphicx}
\usepackage[colorlinks={true}]{hyperref}
\hypersetup{citecolor={blue}, filecolor={blue}, linkcolor={blue}, urlcolor={blue}}

\newtheorem{remark}{Remark}

\IEEEoverridecommandlockouts    

\title{Learning control of quantum systems using frequency-domain optimization algorithms}
\author{Daoyi Dong,
Chuan-Cun Shu, Jiangchao Chen, Xi Xing, Hailan Ma, Yu Guo, Herschel Rabitz
\thanks{This work was supported by the Australian Research
Council's Discovery Projects funding scheme under Project
DP190101566, the National Natural Science Foundation of China
(No. 61828303), NSF (CHE-1464569), ARO (W911NF-16-1-0014), and the U.S. Office of Naval
Research Global under Grant N62909-19-1-2129, and the Alexander von
Humboldt Foundation of Germany.}
\thanks{D. Dong is with the School of Engineering and Information
Technology, University of New South Wales, Canberra, ACT 2600,
Australia, and the Department of Chemistry, Princeton University, Princeton, NJ 08544, USA. (email: daoyidong@gmail.com).}
\thanks{C.C. Shu is with the Hunan Key Laboratory of Super-Microstructure and Ultrafast Process, School of Physics and Electronics, Central South University,
Changsha 410083, China, and the School of Engineering and Information
Technology, University of New South Wales, Canberra, ACT 2600,
Australia.(email:cc.shu@csu.edu.cn)}
\thanks{J. Chen, X. Xing and H. Rabitz are with the Department of Chemistry, Princeton University, Princeton, NJ 08544, USA. (emails: jtc5@princeton.edu, xxing@princeton.edu, hrabitz@princeton.edu).}
\thanks{H. Ma is with the Department of Control and Systems Engineering, School of Management and Engineering, Nanjing University,
Nanjing 210093, China. (email: nju\_mahailan@sina.com).}
\thanks{Y. Guo is with the Hunan Provincial Key Laboratory of Flexible Electronic Materials Genome Engineering, School
of Physics and Electronic Science, Changsha University of Science
and Technology, Changsha 410114, China. (email: guoyu1980@outlook.com).}
}
\begin{document}
\maketitle

\begin{abstract}
We investigate two classes of quantum control problems by using frequency-domain optimization algorithms in the context of ultrafast laser control of quantum systems. In the first class, the system model is known and a frequency-domain gradient-based optimization algorithm is applied to searching for an optimal control field to selectively and robustly manipulate the population transfer in atomic Rubidium. The other class of quantum control problems involves an experimental system with an unknown model. In the case, we introduce a differential evolution algorithm with a mixed strategy to search for optimal control fields and demonstrate the capability in an ultrafast laser control experiment for the fragmentation of Pr(hfac)$_3$ molecules.
\end{abstract}

\section{Introduction}\label{sec:Sec1}
Controlling quantum systems has become an important task in various emerging areas including photophysics, photochemistry, quantum information, and quantum computing \cite{Dong and Petersen 2010}, \cite{Rabitz et al 2000}, \cite{Wiseman and Milburn 2010}, \cite{Nielsen and Chuang 2000}, \cite{GuoShuPRL}, \cite{ShuOL}. A number of control methods including Lyapunov control methodology \cite{Kuang et al 2017}, \cite{KuangTCST}, optimal control theory \cite{Dong and Petersen 2010}, robust control techniques \cite{Dong 2012Automatica,Dong et al 2015,Li et al 2011,Chen et al 2014} and learning control algorithms \cite{Rabitz et al 2000,WuPRA2018,quantum classification} have been proposed to achieve various quantum control goals. Here, we focus on quantum optimal control problems where the goal is to design an optimal control field for a quantum system to achieve a given target. Optimal control theory can be developed to solve this class of quantum control problems. A limitation is that analytical optimal fields can usually be obtained only for low-dimensional quantum systems or simple control tasks. Hence, numerical optimization algorithms find wide applications to search for an optimal (usually suboptimal) field for many quantum control problems \cite{Khaneja et al 2005}. In simulations, the commonly used optimization algorithms are usually performed in the time domain. The application of these algorithms in experiments may become challenging for the use of ultrafast laser pulses with the duration in femtosecond (fs) ($1 \text{fs}=10^{-15}$ second) \cite{Shu2017JPCL} and attosecond (as) ($1 \text{as}=10^{-18}$ second) \cite{ShuJPCL}  regimes, which can be not directly modulated in the time domain. Experimentally, the current pulse shaping technique allows us to shape the temporal field of an ultrafast pulse \cite{Xing NJP} by modulating its spectral phase and/or amplitude in the frequency domain.
This work will demonstrate how frequency-domain optimization algorithms can be employed in numerical simulations and real experiments to search a temporally shaped ultrafast laser pulse for achieving given quantum control tasks.

For the simulations,  we introduce a  frequency-domain optimal control method developed recently in \cite{Shu-PRA2016}, \cite{Shu-PRA2017}, \cite{Guo-PCCP}, which is able to directly calculate the optimized spectral amplitude or phase of an ultrafast laser pulse in the frequency domain while taking account into multiple constraints on the control fields. A gradient-based optimization algorithm is derived for treating a quantum system with known  Hamiltonian. This paper shows how such an optimization algorithm can be utilized to optimize the spectral phase of an ultrafast laser pulse, capable of selectively and robustly controlling quantum state transfer to a desired electronic level in a three-level Rubidium (Rb) atom.

For the experiments, we consider another class of quantum control problems with unknown Hamiltonian (e.g., either for complex quantum systems or when the system is subject to uncertainties). Due to the lack of system model information, it is usually difficult to calculate the gradient of the objective with respect to the control fields, which is crucial in the gradient-based optimization algorithm. To that end, we introduce a frequency-domain differential evolution (DE) algorithm \cite{DE2017} for fragmentation control of Pr(hfac)$_3$ molecules. DE has shown outstanding capability to search for optimal solutions to various complex quantum control problems \cite{Das and Suganthan 2011, Zahedinejad et al 2015}, \cite{Ma 2015SMC}, \cite{Palittapongarnpim et al 2017}.
For example, Zahedinejad and his collaborators \cite{Zahedinejad et al 2015}, \cite{Zahedinejad et al 2016} proposed a subspace-selective self-adaptive DE algorithm for generating high-fidelity quantum gates, showing its high efficiency as compared with the genetic algorithm (GA) and particle swarm optimization (PSO) \cite{Zahedinejad et al 2014}. Recently, Dong and collaborators \cite{DE2017} developed a \textbf{M}ixed-\textbf{S}trategy based DE  algorithm (referred to as MS\_DE) to search for robust control fields in both the time domain and the frequency domain \cite{Ma2017CTT}, \cite{DE2017}. In this paper, we experimentally demonstrate the MS\_DE algorithm for controlling the branching ratio of PrO$^{+}$/PrF$^{+}$ in the photodissociation of Pr(hfac)$_3$ molecules by shaping the spectral phase of ultrafast laser pulses.

The paper is organized as follows. Section \ref{sec:Sec2} provides a detailed introduction to the gradient-based optimization
algorithm in the frequency domain. The application of the gradient-based optimization
algorithm to selective control of quantum state transfer in a three-level
Rb atom is demonstrated in Section \ref{sec:Sec3}. The MS\_DE algorithm is briefly introduced in Section \ref{sec:Sec4}. In Section \ref{sec:Sec5}, we demonstrate experimental results on fragmentation control of Pr(hfac)$_3$ molecules using femtosecond laser pulses. Concluding remarks are given in Section \ref{sec:Sec7}.

\section{Gradient algorithm for learning control of quantum systems}\label{sec:Sec2}
In this section, we assume that the model of a quantum system under consideration is known. Consider an $N$-level quantum system and the dynamical evolution of its state $|\psi(t)\rangle$ can be described by the following Schr\"{o}dinger equation
\begin{equation}
\text{i}\hbar \frac{d}{dt}|\psi(t)\rangle=\hat{H}(t)|\psi(t)\rangle
\end{equation}
where $\text{i}=\sqrt{-1}$, $\hbar$ is the reduced Planck constant, $|\psi(t)\rangle$ is a complex-valued vector (expressed in Dirac notation) in an underlying Hilbert space and $\hat{H}(t)$ is the system Hamiltonian. In the dipole approximation, the system Hamiltonian $\hat{H}(t)$ can be written as
\begin{equation}
\hat{H}(t)=\hat{H}_{0}-\hat{\mu}u(t)
\end{equation}
where $\hat{H}_{0}$ is the free Hamiltonian and $\hat{\mu}$ is the dipole operator. We assume that the eigenvalues of $\hat{H}_{0}$ are $E_{n}$ ($n=1,2,\dots, N$) and the corresponding eigenvectors are $|n\rangle$, $$\hat{H}_{0}=\sum_{n=1}^{N}E_{n}|n\rangle \langle n|$$ where $\langle n|$ is the complex conjugate of $|n\rangle$, i.e., $\langle n|=(|n\rangle)^{\dag}$. The time-dependent evolution of the quantum system from the initial state $|\psi_{0}\rangle$ to $|\psi(t)\rangle$ can be described by a unitary operator $\hat{U}(t)$ with
\begin{equation}
|\psi(t)\rangle=\hat{U}(t)|\psi_{0}\rangle,
\end{equation}
where $\hat{U}^{\dag}\hat{U}=\hat{U}\hat{U}^{\dag}=I$ with identity matrix $I$ and $\hat{U}(0)=I$. The unitary evolution operator $\hat{U}(t)$ is governed by the Schr\"{o}dinger equation
\begin{equation}\label{operator-evolution}
\text{i}\hbar \frac{d}{dt}\hat{U}(t)=\hat{H}(t)\hat{U}(t).
\end{equation}

In the quantum control problem using ultrafast laser fields, the temporal laser field $\mathcal{E}(t)$  can be written as
\begin{equation}
\mathcal{E}(t)=\text{Re}\Bigg[\int_{0}^{\infty}\mathbf{E}(\omega)e^{-\text{i}\omega t}d\omega\Bigg]
\end{equation}
where $\text{Re}(x)$ returns the real part of $x$, the complex spectral field $\mathbf{E}(\omega)$ can be defined in terms of the real-valued spectral amplitude $A(\omega)\geq 0$ and real-valued spectral phase $\phi(\omega)$ as
\begin{equation}
\mathbf{E}(\omega)=A(\omega)e^{-\text{i}\phi(\omega)}.
\end{equation}
The state-of-the-art ultrafast pulse shaping technology has made it possible to manipulate  the spectral amplitude $A(\omega)$
  as well as the spectral phase $\phi(\omega)$ of femtosecond laser pulses. As a result, the temporal control field $\mathcal{E}(t)$ can be designed by shaping the spectral field $\mathbf{E}(\omega)$  in the frequency domain.\\ \indent
 To formulate our method,  the quantum control objective $J$  associated with the expectation value of an arbitrary observable $\hat{O}$ at the end of the control with $0\leq t \leq T$ can be
expressed as
 \begin{equation}\label{obj0}
J(\hat{O})=\mathrm{Tr}\left[
\hat{U}(T)|\psi_{0}\rangle\langle\psi_0|\hat{U}^{\dagger}(T)\hat{O}\right]
\end{equation}
where $\hat{O}$ is a Hermitian operator and $\mathrm{Tr}(A)$ denotes the trace of $A$.
Now we introduce a dummy variable $s\geq 0$ to track the trajectory for optimizing the spectral field $\mathbf{E}(\omega)$. Then the gradient of $J$ with respect to $s$ can be expressed as
\begin{equation}\label{obj}
g_{0}(s)\equiv\frac{dJ}{ds}=\int_{0}^{\infty}\frac{\delta J}{\delta \mathbf{E}(s,\omega)}\frac{\partial \mathbf{E}(s,\omega)}{\partial s}d\omega.
\end{equation}
We aim to develop an iterative algorithm to optimize the objective function $J$. To maximize $J$, we expect $\frac{dJ}{ds}\geq 0$ during the iterative process. The condition can be satisfied in the absence of constraints by choosing
\begin{equation}\label{uncons}
\frac{\partial \mathbf{E}(s,\omega)}{\partial s}=\Bigg[\frac{\delta J}{\delta \mathbf{E}(s,\omega)}\Bigg]^{\ast}
\end{equation}
where $a^{*}$ denotes the conjugate of $a$.

In practical applications, (\ref{uncons}) may be generalized to include  a set of  equality constraints $f_{k}(\mathbf{E}(s,\omega))=C_{k}$, $k=1,\dots,K$. During the optimization process, these constraints can be written as
\begin{equation} \label{obj2}
g_{l}(s)\equiv \frac{df_{k}}{ds}=\int_{0}^{\infty} \frac{\delta f_{k}}{\delta E(s,\omega)}\frac{\partial E(s,\omega)}{\partial s}d\omega=0.
\end{equation}
The combined requirements in (\ref{obj}) and
(\ref{obj2}) can be fulfilled at the same time  by \cite{Guo-PCCP}
\begin{equation}\label{field-gradient}
\frac{\partial E(s,\omega)}{\partial s}=g_{0}(s)\int_{0}^{\infty}S(\omega'-\omega)\sum_{k=0}^{K}[\Gamma^{-1}]_{0k}c_{k}^{\ast}(s,\omega')d\omega',
\end{equation}
where $S(\omega'-\omega)$ is the filter function for smoothing the distribution of the spectral phase \cite{Shu-PRA2016}, $c_{k}(s,\omega)$ is defined by

 \begin{equation} \label{constraints}
c_{k}(s,\omega)=\left\{
                  \begin{array}{ll}
                    \frac{\delta J}{\delta E(s,\omega)}, & k=0; \\
                    \frac{\delta f_{k}}{\delta E(s,\omega)}, & k=1,\dots,K
                  \end{array}                \right.
\end{equation}
and the  elements of the $(K+1)\times (K+1)$ symmetric matrix $\Gamma$ are given by
\begin{equation}
\Gamma_{kk'}=\int_{0}^{\infty}c_{k}(s,\omega)\int_{0}^{\infty}S(\omega'-\omega)c_{k'}^{\ast}(s,\omega')d\omega'd\omega.
\end{equation}

In practical implementation, we can separately calculate the gradients of $J$ with respect to $A(s,\omega)$ and $\phi(s,\omega)$. This in turn leads to two common used control experimental schemes, i.e., the spectral amplitude control and the spectral phase-only control. For numerical simulations, the two control schemes can be described by
\begin{equation}
\frac{\delta J}{\delta A(s,\omega)}=\int_{-\infty}^{\infty}\frac{\delta J}{\delta \mathcal{E}(s,t)}\frac{\partial \mathcal{E}(s,t)}{\partial A(s,\omega)}dt,
\end{equation}
\begin{equation}
\frac{\delta J}{\delta \phi(s,\omega)}=\int_{-\infty}^{\infty}\frac{\delta J}{\delta \mathcal{E}(s,t)}\frac{\partial \mathcal{E}(s,t)}{\partial \phi(s,\omega)}dt.
\end{equation}
The gradients $\frac{\partial \mathcal{E}(s,t)}{\partial A(s,\omega)}$ and $\frac{\partial \mathcal{E}(s,t)}{\partial \phi(s,\omega)}$ in our situation are analytically given by
\begin{equation}
\frac{\partial \mathcal{E}(s,t)}{\partial A(s,\omega)}=\cos[\phi(s,\omega)-i\omega t],
\end{equation}
\begin{equation}
\frac{\partial \mathcal{E}(s,t)}{\partial \phi(s,\omega)}=-A(\omega)\sin[\phi(s,\omega)-i\omega t],
\end{equation}
and the gradient $\frac{\delta J}{\delta \mathcal{E}(s,t)}$ can be expressed as \cite{Shu-PRA2016}
\begin{equation}
\frac{\delta J}{\delta \mathcal{E}(s,t)}=-2\text{Im}(\text{Tr}\{[\psi_{0}\rangle\langle\psi_0,\hat{U}^{\dagger}\hat{O}\hat{U}(T)]\hat{U}^{\dagger}\hat{\mu}\hat{U}(T)\})
\end{equation}
where $\text{Im}(x)$ returns the imaginary part of $x$, and $[A, B]=AB-BA$.

\begin{remark}
In the above process of obtaining the gradient algorithm for optimizing an objective function, we assume that the system model is known and the system evolution is described by a Schr\"{o}dinger equation. The algorithm is applicable for a finite dimensional closed quantum system or a quantum system that can be approximated as a finite dimensional closed system. In many practical applications, a quantum system under consideration may need to be described as an open quantum system. Then the system state needs to be described by a density operator $\rho$ satisfying $\rho \geq 0$, $\rho=\rho^{\dag}$ and $\text{Tr}(\rho)=1$. If we know the system model (e.g., a Markovian master equation for a Markovian open quantum system \cite{Dong and Petersen 2010}), we can also develop a gradient algorithm to find an optimal (usually suboptimal) solution to an optimal control problem of the quantum system. During the development of such an algorithm, the difference from the case of closed quantum systems is that the system evolution from one state to another state needs to be described by a superoperator instead of a unitary transformation \cite{Wu et al 2017}.
\end{remark}

\section{Numerical results  on selective control of atomic Rb}\label{sec:Sec3}
\begin{figure}[!t]\centering
\resizebox{0.8\textwidth}{!}{%
  \includegraphics{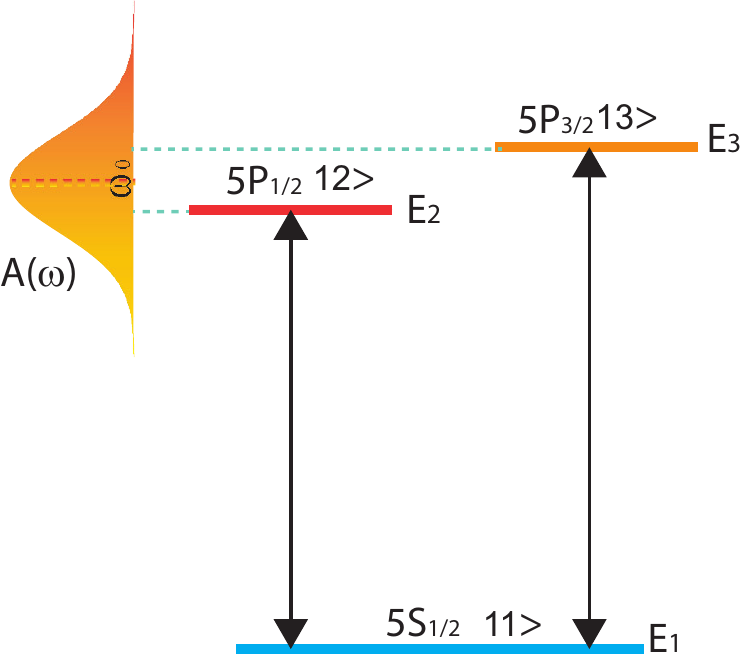}
} \caption{Spectral phase only control scheme in a three-level V-type Rb atom.
 The laser pulse with a fixed spectral amplitude $A(\omega)$ is used to excite Rb atoms from the ground electronic state $5S_{1/2}$ to  excited electronic states $5P_{1/2}$ and $5P_{3/2}$, whose branching ratio is controlled by optimizing the spectral phase $\phi(\omega)$ of the laser pulse. In our simulations, $5S_{1/2}$, $5P_{1/2}$ and $5P_{3/2}$ are denoted by three states $|1\rangle$, $|2\rangle$ and $|3\rangle$ with energies $E_1$, $E_2$ and $E_3$, respectively.
 } \label{fig1}
\end{figure}

To illustrate the frequency-domain optimization algorithm in Section \ref{sec:Sec2}, we consider a three-level V-type system in Fig. \ref{fig1}, which consists of  the ground electronic state  $5S_{1/2}$ and the two lowest excited electronic states $5P_{1/2}$ and $5P_{3/2}$ of a Rubidium atom, denoting  by $|1\rangle$, $|2\rangle$, and $|3\rangle$ with energies $E_1=0$, $E_2=12578.95$ cm$^{-1}$, $E_3=12816.55$ cm$^{-1}$, respectively. The free Hamiltonian $\hat{H}_0$ is given by
 \begin{eqnarray}
\hat{H}_0=\left(
\begin{array}{cccccccc}
 E_1 &0 &0 \\  0 & E_2 & 0 \\ 0 &0 & E_3
\end{array} \right).
\end{eqnarray}
We fix the spectral amplitude of the ultrafast laser pulse unchanged with a Gaussian distribution  $$A(\omega)=\mathcal{E}_0\frac{1}{\sqrt{2\pi}\Delta\omega}\exp(-(\omega-\omega_0)^2/2\Delta^2\omega)$$ with $\mathcal{E}_0=3.6\times10^6$ V/cm,   $\omega_0=(E_2+E_3)/2\hbar=12698$ cm$^{-1}$ and $\Delta\omega=177$ cm$^{-1}$ to excite this three-level system.  The dipole moment operator $\hat{\mu}$ is given by
\begin{eqnarray}
\label{dipole}\hat{\mu}=\left(
\begin{array}{cccccccc}
 0 & \mu_{12} & \mu_{13} \\  \mu_{12} & 0 & 0 \\ \mu_{13} &0 & 0
\end{array} \right),
\end{eqnarray}
with  $\mu_{12}=2.9931$ a.u.  and $\mu_{13}=4.2275$ a.u. \cite{rb-data}, in which the zero matrix elements imply that the transitions between $|2\rangle$ and $|3\rangle$ are forbidden.\\ \indent

We consider two end-point equality constraints
\begin{eqnarray}\label{endconstraint1}
\mathcal{E}(0)=\frac{1}{\sqrt{2\pi}}\int_{-\infty}^{\infty}A(\omega)\cos\left[\phi(\omega) \right]d\omega=0
\end{eqnarray}
and
\begin{eqnarray} \label{endconstraint2}
\mathcal{E}(T)=\frac{1}{\sqrt{2\pi}}\int_{-\infty}^{\infty}A(\omega)\cos\left[\phi(\omega)-\omega T \right]d\omega=0
\end{eqnarray}
on the control field $\mathcal E(t)$ , which enforce that the optimized field is turned on  at $t=0$ and off at $t=T$.  From (\ref{constraints}), we derive the coefficients
$$c_1(s,\omega)=-\frac{1}{\sqrt{2\pi}}A(\omega)\sin\left[\phi(s,\omega)\right]$$
and
$$c_2(s,\omega)=- \frac{1}{\sqrt{2\pi}}A(\omega)\sin\left[\phi(s,\omega)+\omega
T\right].$$ Furthermore, we perform an optimization procedure by shaping the spectral phase of the laser pulse while fixing the spectral amplitude. The optimization algorithm is listed in Algorithm \ref{Gradient-Algorithm}.
\begin{algorithm}
\caption{Algorithmic description of gradient algorithm}\label{Gradient-Algorithm}
\begin{algorithmic}[1]
\State Solve (\ref{operator-evolution}) using an initial field $\mathcal{E}(s_{0},t)$ with a guess of the spectral phase $\phi(s_{0},\omega)$;
\State  Calculate $\frac{\delta \textbf{J}}{\delta \phi(s_{0},\omega)}$;
\State Solve the first-order differential equation (\ref{field-gradient}) to obtain the first updated spectral phase $\phi(s_{1}=s_{0}+\delta s, \omega)=\phi(s_{0},\omega)+\delta s \frac{\partial \phi(s_{0},\omega)}{\partial s}$;
\State Repeat Step 1 through Step 3 with the updated spectral phase as the initial guess until the ``best" spectral phase is found.
\end{algorithmic}
\end{algorithm}

\begin{figure}[!t]\centering
\resizebox{1.0\textwidth}{!}{%
  \includegraphics{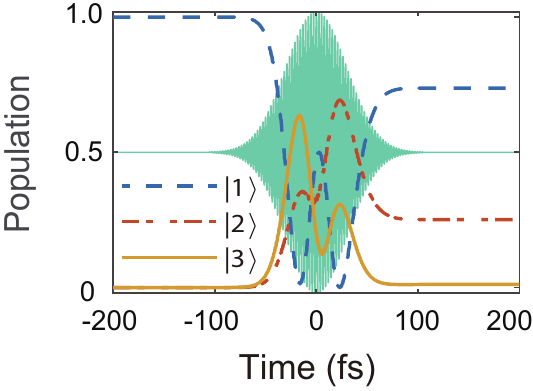}
} \caption{The time-dependent population transfer among
three states $|1\rangle$, $|2\rangle$, and $|3\rangle$  with a transform-limited pulse in green.} \label{fig2}
\end{figure}

We first consider a flat spectral phase of $\phi(\omega)$ and fixed spectral amplitude $A(\omega)$ at zero, which corresponds to a transform-limited pulse $$\mathcal{E}(s_{0},t)=\mathcal{E}_0
\exp(-t^2/2\tau_0^2)\cos\omega_0t$$ with a duration of $\tau_0=1/\Delta\omega=30$ fs.
Figure \ref{fig2} shows the time-dependent evolution of the quantum state transfer among the three levels. Some oscillations between levels in the population transfer take place. After the pulse is off at $T=200$ fs, all three levels are populated. In the following simulations, we selectively maximize quantum state transfer to either state $|2\rangle$ or state $|3\rangle$  by optimizing  the spectral phase $\phi(\omega)$ of the laser pulse under two end-point equality constraints by (\ref{endconstraint1}) and (\ref{endconstraint2}) while keeping its spectral amplitude $A(\omega)$ unchanged. \\ \indent
\begin{figure}[!t]\centering
\resizebox{1.0\textwidth}{!}{%
  \includegraphics{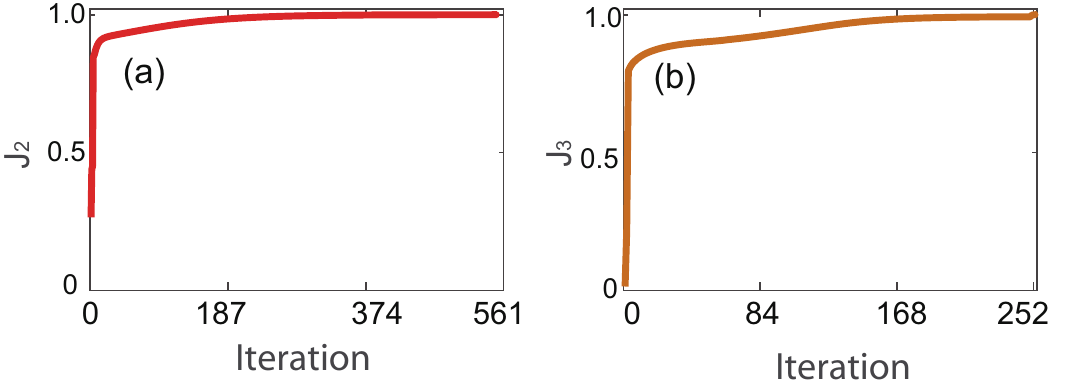}
} \caption{The control objectives $J_2=|\langle2|\psi(T)|^2$ and $J_3=|\langle3|\psi(T)|^2$ vs iterations with $\sigma=50$$^{-1}$.} \label{fig3}
\end{figure}
\begin{figure}[!t]\centering
\resizebox{1.0\textwidth}{!}{%
  \includegraphics{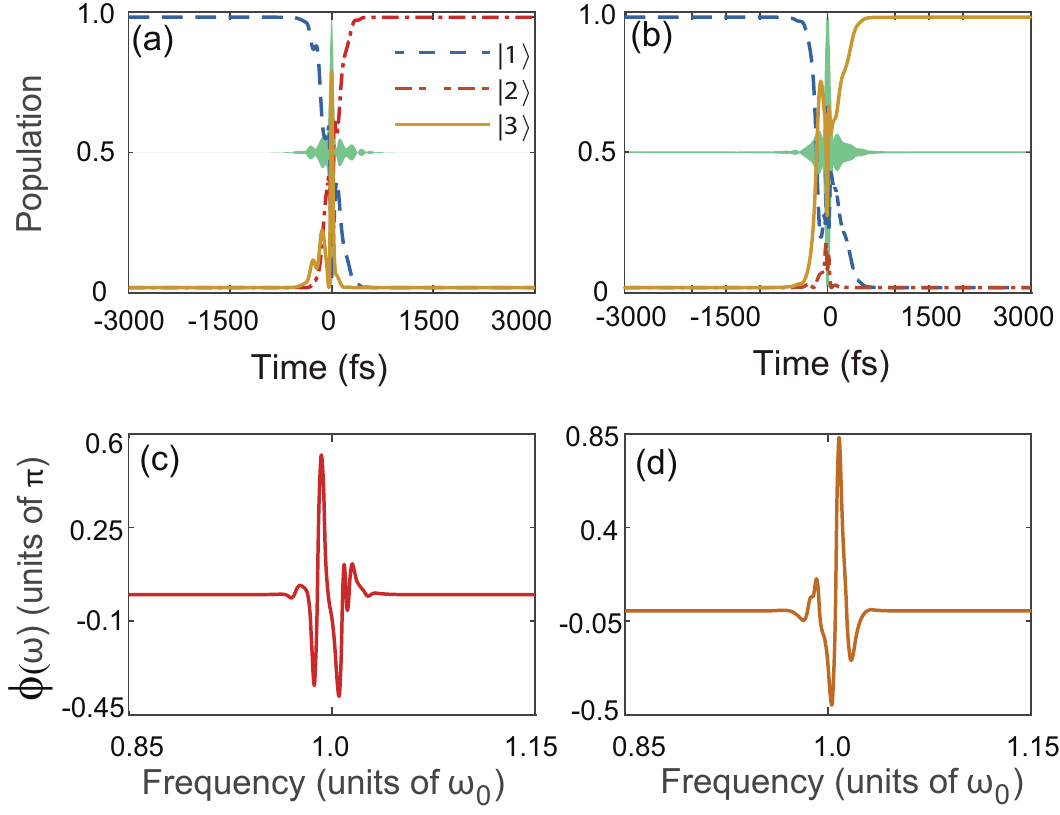}
} \caption{Optimal population evolutions and spectral phases with $\sigma=50$ cm$^{-1}$. The population evolutions of quantum state transfer to the target state (a) $|2\rangle$ and (b) $|3\rangle$ with laser pulses in green. The corresponding optimized spectral phases for (a) and (b) are shown in (c) and (d).} \label{fig4}
\end{figure}

\begin{figure}[!t]\centering
\resizebox{1.0\textwidth}{!}{%
  \includegraphics{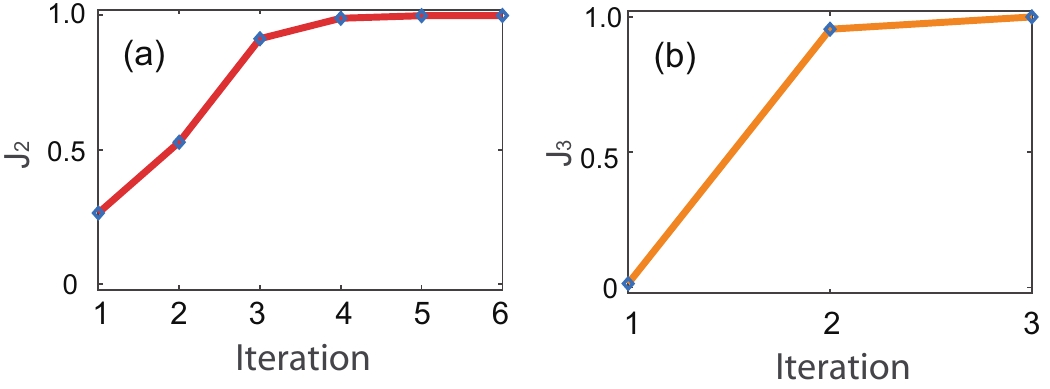}
} \caption{The control objectives (a) $J_2=|\langle2|\psi(T)|^2$ and (b) $J_3=|\langle3|\psi(T)|^2$ vs iterations with $\sigma=5000$ cm$^{-1}$.} \label{fig5}
\end{figure}

\begin{figure}[!t]\centering
\resizebox{1.0\textwidth}{!}{%
  \includegraphics{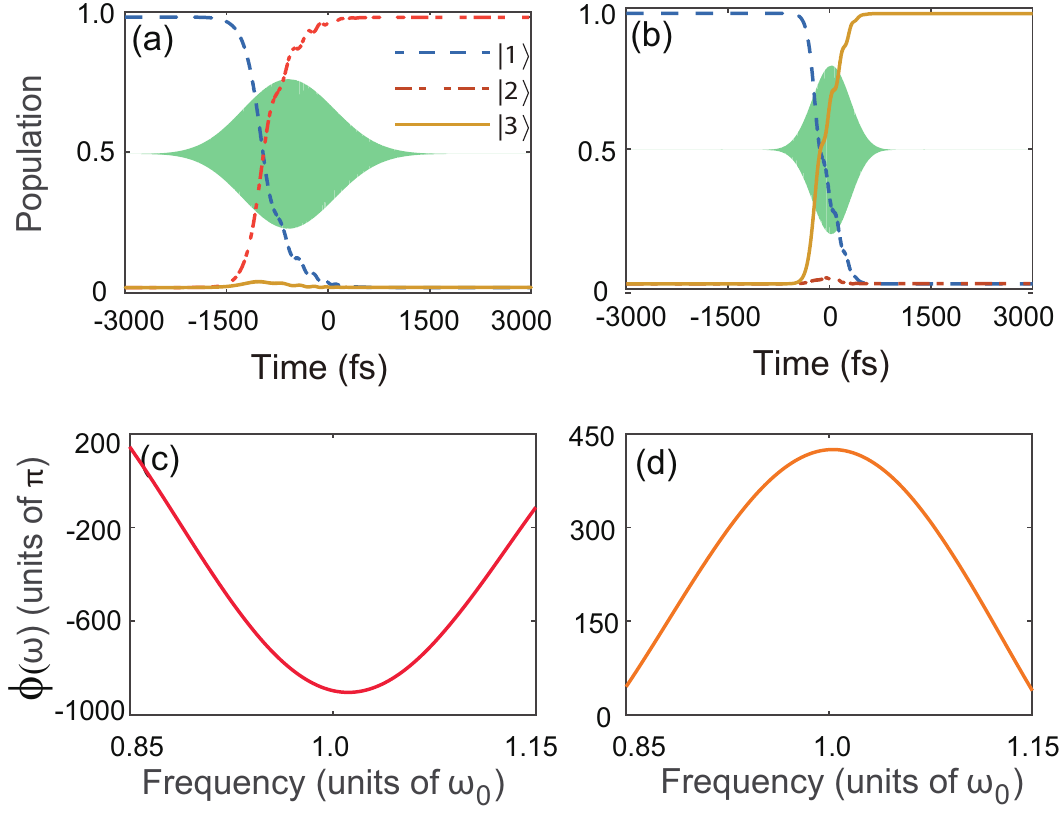}
} \caption{Optimal population evolutions and spectral phases with $\sigma=5000$$^{-1}$. The population evolutions of quantum state transfer to the target state (a) $|2\rangle$ and (b) $|3\rangle$. The corresponding optimized spectral phases for (a) and (b) are shown in (c) and (d), respectively.} \label{fig6}
\end{figure}

To achieve the goal, we define the observable $O=|j\rangle\langle j|$ with $j=2$ or 3  to maximize quantum state transfer to $|2\rangle$ or $|3\rangle$, respectively. We start with $\mathcal{E}(s_{0},t)$ as the initial guess, and take a normalized Gaussian function of the form $$S(\omega'-\omega)=\exp(-4\ln2(\omega-\omega')^2/\sigma^2)$$ with a filter parameter $\sigma$. Since the speed of convergence and the shape of the optimized spectral phase are highly dependent on the choice of stepsize $\delta s$ and  the parameter of $\sigma$ in the filter function $S$, we examine two different cases with a small value of $\sigma=50$ cm$^{-1}$ and a large value of $\sigma=5000$ cm$^{-1}$ and $ds$ is varied adaptively during iterations.  Figure \ref{fig3} shows the control objectives $J_2=|\langle2|\psi(T)|^2$ and $J_3=|\langle3|\psi(T)|^2$ as a function of iterations with a small value of  $\sigma=50$ cm$^{-1}$. After a few hundred iterations, both objectives can monotonically increase to a very high fidelity of $>0.9999$ by only optimizing the spectral phase.  As a result, it is possible to selectively control population transfer to the excited electronic states $|2\rangle$ and $|3\rangle$. Fig. \ref{fig4} shows the time-dependent populations with the optimized control fields and the corresponding optimized spectral phases. We can see that the populations are successfully transferred from the initial state to the target state, whereas another state is populated significantly during the transfer process, as shown in Fig. \ref{fig4} (a) and (b). The optimized spectral phases in Fig. \ref{fig4} (c) and (d) are complex and exhibit strong oscillations in the frequency domain. The solutions to obtaining a high fidelity quantum state transfer in such a simple quantum system are not unique. If we further decrease the value of $\sigma$ in the filter $S$, the optimized spectral phases will become more complex than that in Fig. \ref{fig4} (c) and (d).
 \\ \indent
We now examine the optimization algorithm with a large value of $\sigma=5000$ cm$^{-1}$ and demonstrate the corresponding results in Fig. \ref{fig5}. It is surprising that the control objectives can rapidly and monotonically increase to a high fidelity of $>0.999$  after a few iterations.  Figure \ref{fig6} shows the time-dependent population dynamics with the optimized fields and the corresponding optimized spectral phases. The population dynamics change significantly with the optimized fields as compared with that in Fig. \ref{fig4}. It is also interesting to note that the populations are successfully transferred from the initial state $|1\rangle$ to the target state $|2\rangle$ in Fig. \ref{fig6} (a)  and $|3\rangle$ in Fig. \ref{fig6} (b), while the population transfer to another excited electronic state beyond the target state is clearly suppressed during the quantum state transfer process. \\ \indent

The optimized spectral phases in Figs. \ref{fig6} (c) and (d) with constant shifts can be fitted very well by using a quadratic function of $\beta_0(\omega-\omega_c)^2$ with a chirp rate $\beta_0$ and a modulated frequency $\omega_c$.
To that end, the optimized spectral phases are found   with $\beta_0=7191$ fs$^2$ and $\omega_c=1.28423\times10^4$ cm$^{-1}$ in Fig. \ref{fig6} (c), and with $\beta_0=-3018$ fs$^2$ and $\omega_c=1.27329\times10^4$ cm$^{-1}$ in Fig. \ref{fig6} (d), which significantly prolong the optimized fields in the time domain while reducing  their amplitudes as compared with the initial transform-limited pulse.  A positively chirped pulse in Fig. \ref{fig6} (a) maximizes population transfer to state $|2\rangle$ by suppressing the transfer to state $|3\rangle$, whereas a negatively chirped pulse in Fig. \ref{fig6} (b)  leads to high efficiency of population transfer to state $|3\rangle$ by reducing the transfer to state $|2\rangle$. It implies that an adiabatic passage is constructed between the initial and target states, and therefore robust quantum state transfer is obtained against the variations of the system and control field. To that end, our method can provide a new approach to search a robust control field by shaping the spectral phase of the laser pulse in the frequency domain.

\begin{remark}
In the numerical example, we consider the phase-only control of atomic Rb that can be approximately considered as a three-level quantum system. The algorithm is also applicable for other finite-level quantum systems using phase-only control or amplitude-only control as long as we know the system model so that the gradient of a given objective function $J$ with respect to relevant control variables can be derived in an analytical way. For many practical applications, reliable quantum system models may be unknown. In such a situation, it is not convenient to directly calculate the gradient required for the optimization algorithms. A possible strategy is that we may first identify the system model and then employ a gradient iterative algorithm to find an optimal control field. A number of identification methods has been developed to identify the reaction mechanism, system dimension or system Hamiltonian for various quantum systems \cite{Shu2017JPCL}, \cite{WangTAC2018}, \cite{burgarth 2012}, \cite{sone dimension}, \cite{WangAutomatica}.
However, for more complex quantum system or quantum process, it is usually difficult to identify the dynamic model before controlling it. Instead, we may employ closed-loop quantum control scheme to learn optimal ultrafast laser pulses in the frequency domain for controlling quantum systems. In the following, we will employ a DE algorithm to experimentally investigate ultrafast laser control of complex molecules Pr(hfac)$_3$.
\end{remark}

\section{Differential evolution for learning control of quantum systems with unknown models}\label{sec:Sec4}
Among evolutionary computing algorithms, differential evolution (DE) is a simple but powerful stochastic search technique. It has been widely used in the continuous search domain and has achieved considerable success in science and engineering applications \cite{Das and Suganthan 2011}, \cite{Storn and Price 1997}, \cite{Storn and Price 1995}, \cite{Sarker et al 2016}. In DE, a population is composed of a group of individual trial solutions or \emph{parameter vectors}, usually represented in a real-valued vector $ X=[x_1, x_2, \cdots, x_D ]^T$. In the process of searching, an objective function regarding a target vector $X$ is defined as $J(X)$. Then, the learning process works by generating variations of the individuals within the given parameter space and selecting the better to be carried into the next generation, until a ``best" individual is obtained.
Consider that DE searches for a global optimum point in a $D$-dimensional real parameter space $\Re^D$.  We can summarize its main steps as follows.

(a) \textbf{Initialization}.
 The population (i.e., target vector) at the current generation is denoted as $X_{i, G}=[{x_{i, G}^1, \cdots, x_{i, G}^D}]^{T}$, $i=1, ..., NP$ and let  $x_{i,G}^j \in[x_{\min}^j,x_{\max}^j]$, $(j=1,2,...,D)$. Usually, the population (at $G=0$) are initialized in a uniform way \cite{Mallipeddi et 2011}:
 \begin{equation}\label{eq:initialization}
  x_{i,0}^j=x_{\min}^j+\textup{rand}(0,1)\cdot(x_{\max}^j-x_{\min}^j), \quad j=1,2,...,D,
 \end{equation}
where $\textup{rand}(0,1)$ is a uniformly distributed random number, which helps guarantee that the vectors cover the range of the parameter space.

(b) \textbf{Mutation}. The core idea of the ``mutation" operation is to generate mutant vectors from the existing target vectors. For example, by choosing three other distinct parameter vectors from the current generation (say, $X_{r_1}$, $X_{r_2}$, $X_{r_3}$), we could formulate a donor vector $V_{i,G+1}$ as
    \begin{equation}\label{eq:mutation}
    V_{i, G}=X_{r_1, G}+F \cdot(X_{r_2,G}-X_{r_3, G}),
    \end{equation}
where the indices $r_1, r_2, r_3 \in \{1, ..., NP\}$ are mutually exclusive integers from $[1,NP]$ and $r_1, r_2, r_3\neq i$. Besides, the scaling factor $F$ is normally set between 0.4 and 1.

(c) \textbf{Crossover}. In DE, a mutation phase is usually followed by a crossover operation, as it generate trial vectors from mutant vector $V_{i, G}$ and target vector $X_{i, G}$. There are two typical crossover operations including exponential crossover and binomial crossover. They are functionally equivalent to each other, and the binomial one could be expressed as:
\begin{equation}\label{eq:crossover}
    u_{i, G}^j=\left \{
\begin{split} \displaystyle
    v_{i, G}^j, \ \ \ \text{if}\ \text{rand}(j) \leq CR \ \text{or} \ j=\textup{rand}(1, D), \\
    x_{i, G}^j, \ \ \text{if} \ \text{rand}(j)>CR\ \text{and} \ j \neq \textup{rand}(1, D),\\
\end{split}\right.
\end{equation}
where the pre-defined parameter $CR$ controls the potential diversity of the evolving population. The condition $j=\textup{rand}(1, D) $ is introduced to ensure that the trial vector is different from its corresponding target vector by at least one parameter.

(d) \textbf{Selection}. After the crossover, a selection process is adopted to determine the individuals of the next generation from the target vectors and the trial vectors by comparing their fitness functions. For a maximization problem, if the new trial vector yields an equal or higher value of the objective function, it survives into the next generation; otherwise the target vector retains. This could be outlined as:
    \begin{equation}\label{eq:selection}
    X_{i, G+1}=\left \{
\begin{split} \displaystyle
    U_{i, G}, \ \text{if} \quad J(U_{i, G})\geq J(X_{i, G}), \\
    X_{i, G}, \ \text{otherwise}. \ \ \ \ \ \ \ \ \ \ \ \ \
\end{split}\right.
    \end{equation}



\begin{algorithm}
\caption{Algorithmic description of \emph{MS}\_DE}\label{Algorithm EMSDE}
\begin{algorithmic}[1]
\State Set the generation number $G=0$
\For {$i=1$ to $NP$}  
\State  initialize $ X_{i,0}$ and obtain fitness function $J(X_{i,0})$
\EndFor
\State {Set the vector with the maximum fitness as $X_{\text{best},0}$}

\Repeat {\ (for each generation $G=0,1,\ldots,G_{\max}$)}
\Repeat {\ (for each vector $X_i$, $i=1,2,\ldots,NP$)}

\State Set parameter $F_{i,G}=\text{Normrnd}(0.5,0.3)$
\State Set parameter $CR_{i,G}=\text{Normrnd}(0.5,0.1)$
\While {$CR_{i,G}<0$ or $CR_{i,G}>1$}
\State $CR_{i,G}=\text{Normrnd}(0.5,0.1)$\
\EndWhile

\State {randomly choose a \textbf{strategy} from candidate pool}
\State  {obtain mutant vectors $V_{i,G}$ according to (\ref{eq:strategy 1})-(\ref{eq:strategy 4}) }

%
%
%

\If  {$\textbf{stragegy} \in \{1,2,3\}$}
\State {obtain $U_{i,G}$ according to equation ({\ref{eq:crossover}}) }
\EndIf
\If  {$\textbf{strategy} \in \{4\}}$ $U_{i,G}=V_{i,G}$
\EndIf


\If  {$J(U_{i,G}) \ge J(X_{i,G})$} \label{algo:renew begin}
\State  $X_{i,G+1} \leftarrow U_{i,G}$,\quad $J(X_{i,G+1}) \leftarrow J(U_{i,G})$.
\EndIf \label{algo:renew end}

\State {Update the best vector $X_{\text{best},G}$ and $i \leftarrow i+1$}
\Until {\ $i=NP$}

\State  $G \leftarrow G+1$
\Until {\ $G=G_{\max}$}
\end{algorithmic}
\end{algorithm}

According previous studies \cite{Storn and Price 1995}, \cite{Qin et al 2009}, mutation operation aims at generating variations for the population, therefore the adopted mutation strategy can have key impact on its searching performance. Existing results have shown that different mutation strategies exhibit various optimization effects on different searching problems and they are suitable for solving different specific optimization problems \cite{Das and Suganthan 2011}, \cite{Neri and Tirronen 2010}, \cite{Becerra and Coello 2006}. In \cite{Ma2017CTT}, a DE algorithm with mixed strategies has been demonstrated to be a promising candidate for quantum control problems. In this paper, we adopt the DE algorithm with a mixed strategy (i.e., MS\_DE algorithm in Algorithm \ref{Algorithm EMSDE}) for solving the quantum control problem with an unknown model in the frequency domain.
Here, we denote the mutation strategy using the notation
DE/x/y, where $x$ represents a string denoting the
base vector to be perturbed, $y$ is the number of difference
vectors considered for perturbation of $x$. To guarantee the optimization effects, we investigate several strategies and finally decide on four effective yet diverse strategy candidates \cite{Ma2017CTT}, \cite{Becerra and Coello 2006}. They are as follows:

%

\textbf{strategy 1}: DE/rand/1
\begin{equation}\label{eq:strategy 1}
    V_i=X_{r_1}+F\cdot(X_{r_2}-X_{r_3}).
\end{equation}

\textbf{strategy 2}: DE/rand to best/2
\begin{equation}\label{eq:strategy 2}
    V_i=X_i+F\cdot(X_{best}-X_i)+F\cdot(X_{r_1}-X_{r_2})+F\cdot(X_{r_3}-X_{r_4}).
\end{equation}

\textbf{strategy 3}: DE/rand/2
\begin{equation}\label{eq:strategy 3}
    V_i=X_{r_1}+F\cdot(X_{r_2}-X_{r_3})+F\cdot(X_{r_4}-X_{r_5}).
\end{equation}

\textbf{strategy 4}: DE/current-to-rand/1
\begin{equation}\label{eq:strategy 4}
    V_i=X_i+K\cdot(X_{r_1}-X_i)+F\cdot(X_{r_2}-X_{r_3}).
\end{equation}
The indices $r_1, r_2, r_3, r_4$ and $r_5$ are mutually exclusive integers randomly chosen from the range $[1,NP]$ and all of them are different from the index $i$. $X_\textup{best}$ is the best individual vector with the best fitness (i.e., the highest objective function value for a maximization problem) in the population.
To eliminate one additional parameter, the control parameter $K$ in ({\ref{eq:strategy 4}}) could be set as $K=0.5$.
In the MS\_DE algorithm, a mutation scheme from a candidate pool is selected and then crossover operation is determined. In our implementation, the first three mutation schemes are combined with a binomial crossover operation, while the fourth scheme directly generates trial vectors without crossover.

As for control parameters of DE, we sample $F$ by  a normal distribution with mean value 0.5 and standard deviation 0.3, denoted by $\text{N}(0.5,0.3)$. Similarly, the value of $CR$ is sampled by a normal distribution denoted as $\text{N}(0.5,0.1)$. Considering that $CR$ has the practical meaning of probability, those values falling out $[0,1]$ should be abandoned and new values should be regenerated. Consequently, a set of $F$ and $CR$ values are assigned to each target vector for performing mutation, crossover and selection.

\section{Experimental results on fragmentation control of Pr(hfac)$_3$ using femtosecond laser pules}\label{sec:Sec5}

\subsection{Pr(hfac)$_3$ molecule}
\begin{figure}
\centering
\includegraphics[width=0.7\textwidth]{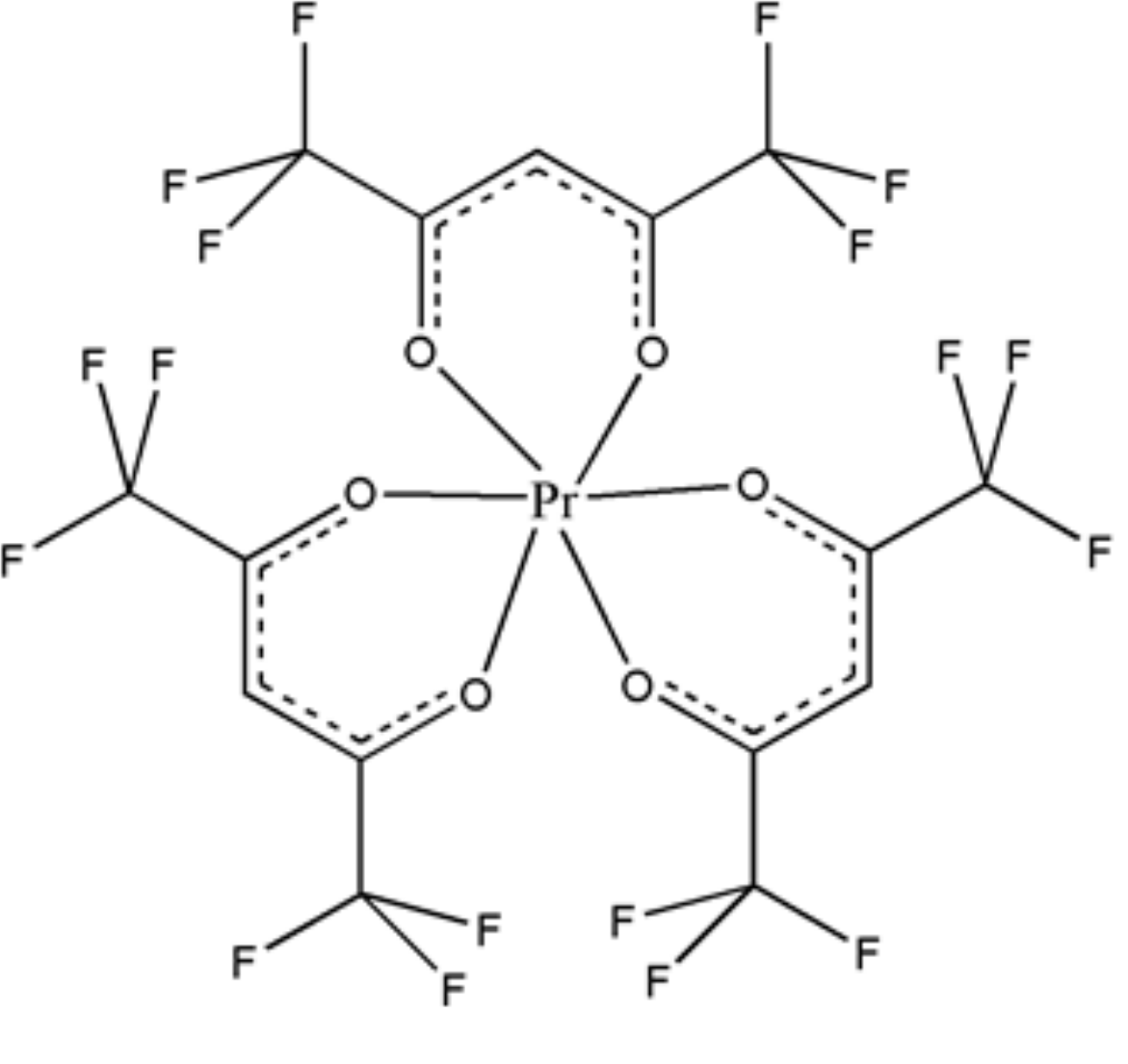}
\caption{Structure of a Pr(hfac)$_3$ molecule.}
\label{molecule}
\end{figure}

Fluorinated praseodymium complexes Pr(hfac)$_3$ (hfac = hexafluoroacetylacetonate) molecules are a common precursor for making thin films of praseodymium materials with metal-organic chemical vapor deposition, because of their high thermal stability and volatility \cite{Chen1,Chen2} and superior transport properties \cite{Chen3,Chen4}. The molecular structure of Pr(hfac)$_3$ is shown in Fig. \ref{molecule}. Even though Pr(hfac)$_3$ is an oxygen-coordinated complex, the praseodymium oxides are not easy to observe using Pr(hfac)$_3$ as a precursor in prior laser-dissociation experiments. Very small amounts of oxide fragments from Pr(hfac)$_3$ were previously reported with continuous-wave and nanosecond lasers \cite{Chen5}. However, Pr(hfac)$_3$ is still an excellent candidate for deposition of praseodymium fluorides \cite{Chen4,Chen6}. The formation of fluorides was explained by Talaga \emph{et al.} \cite{Chen7}, where they proposed a unimolecular reaction that was initiated by rotation of the $C_{\alpha}-C(O)$ bond bringing the CF$_3$ group into proximity to the metal.

Using intense and ultrashort femtosecond laser pulses, it is possible to observe a strong PrO$^{+}$ peak with the precursor Pr(hfac)$_3$. The shaped laser pulses on the fs timescale greatly restrict the $C_{\alpha}-C(O)$ bond rotation and enhance PrO$^{+}$ generation. The results explain why PrO$^{+}$ was rarely observed under continuous-wave and nanosecond laser beams in previous studies.
The purity of the thin praseodymium oxides film and the efficiency to generate oxides are two interesting and valuable problems. Finding the best shaped pulses to optimize the PrO$^{+}$/PrF$^{+}$ ratio is a challenging task. Since we do not know the system model to describe the chemical reaction of Pr(hfac)$_3$ molecules with fs laser pulses, we employ the \emph{MS}\_DE in Section \ref{sec:Sec4} to find an optimal field to control the PrO$^{+}$/PrF$^{+}$ fragmentation ratio in Pr(hfac)$_3$ molecules.

\begin{figure*}
\centering
\includegraphics[width=0.80\textwidth]{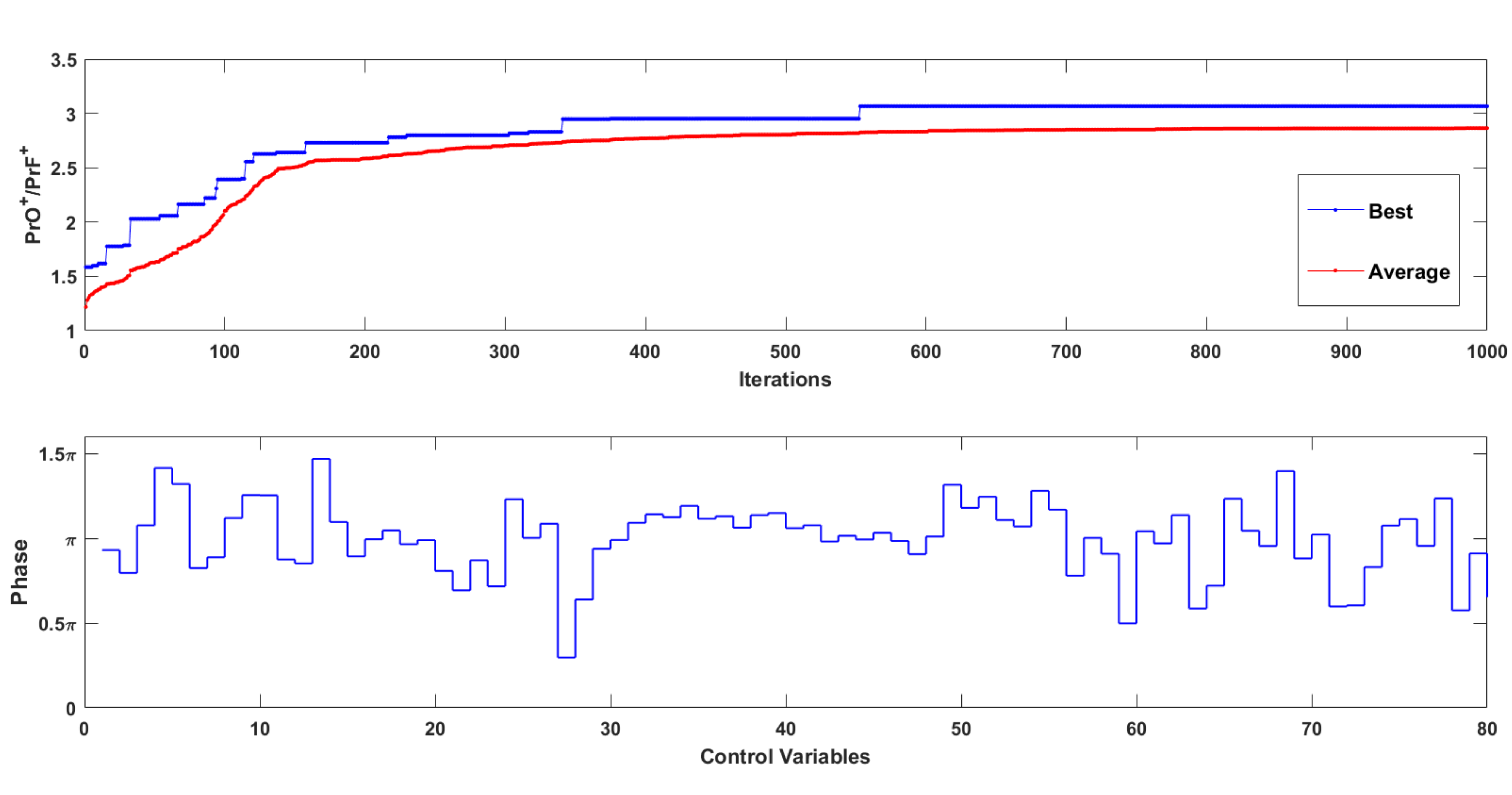}
\caption{Experimental result on the femtosecond laser control system for optimizing the ratio between the products PrO$^{+}$ and PrF$^{+}$ using MS\_DE when no constraint on the phase. (a) Ratio PrO$^{+}$/PrF$^{+}$ vs iterations, where `Best' represents the maximum fitness and `Average' represents the average fitness of all individuals during each iteration. (b) Optimized phases of 80 control variables for the optimal result that corresponds to the maximum fitness.}
\label{Experiment2pi}
\end{figure*}

\begin{figure*}
\centering
\includegraphics[width=0.80\textwidth]{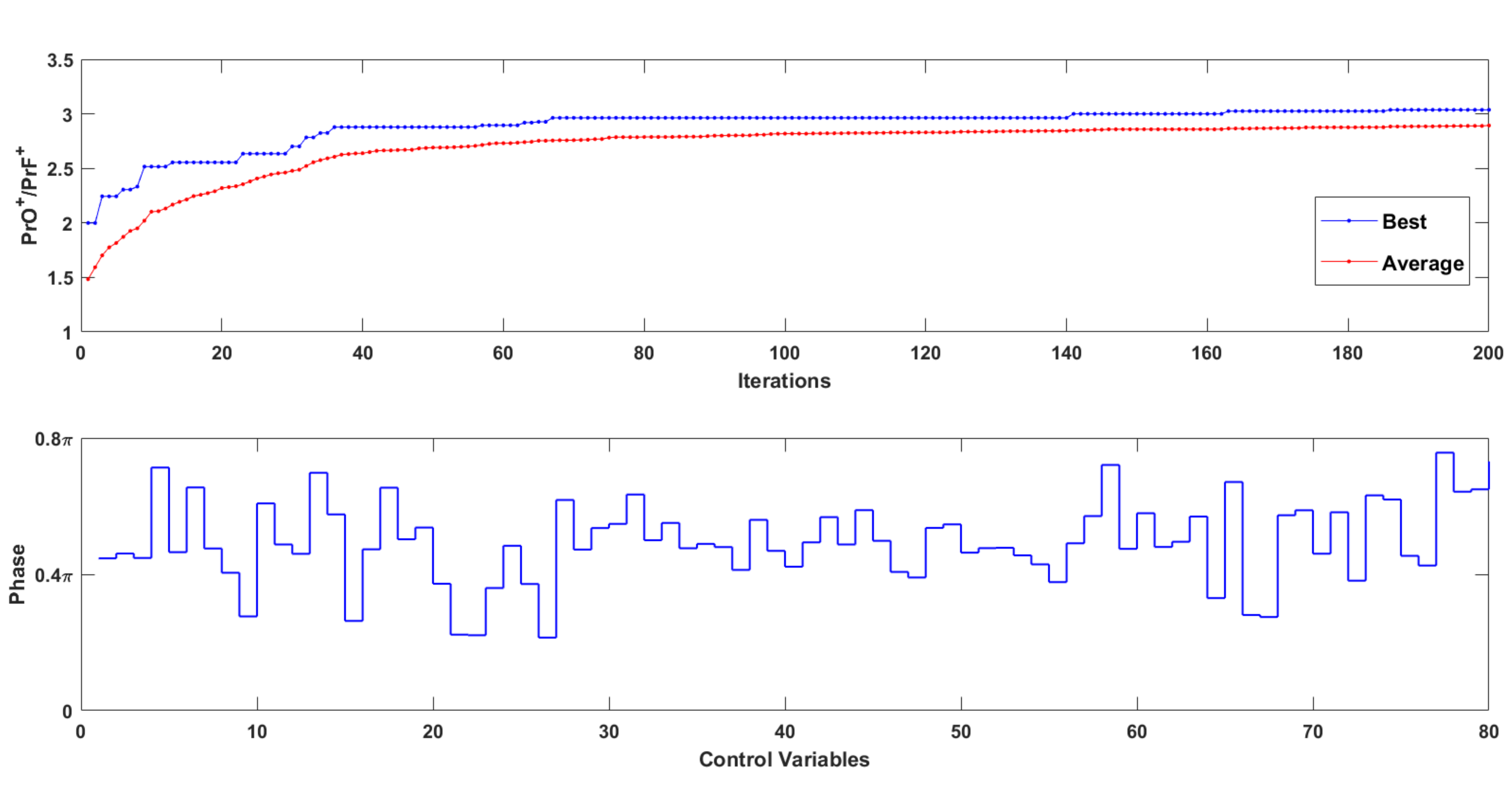}
\caption{Experimental result on the femtosecond laser control system for optimizing the ratio between the products PrO$^{+}$ and PrF$^{+}$ using MS\_DE when the phase is constrained in $[0, \pi]$. (a) Ratio PrO$^{+}$/PrF$^{+}$ vs iterations, where `Best' represents the maximum fitness and `Average' represents the average fitness of all individuals during each iteration. (b) Optimized phases of 80 control variables for the optimal result that corresponds to the maximum fitness.}
\label{Experimentpi}
\end{figure*}

\begin{figure*}
\centering
\includegraphics[width=0.80\textwidth]{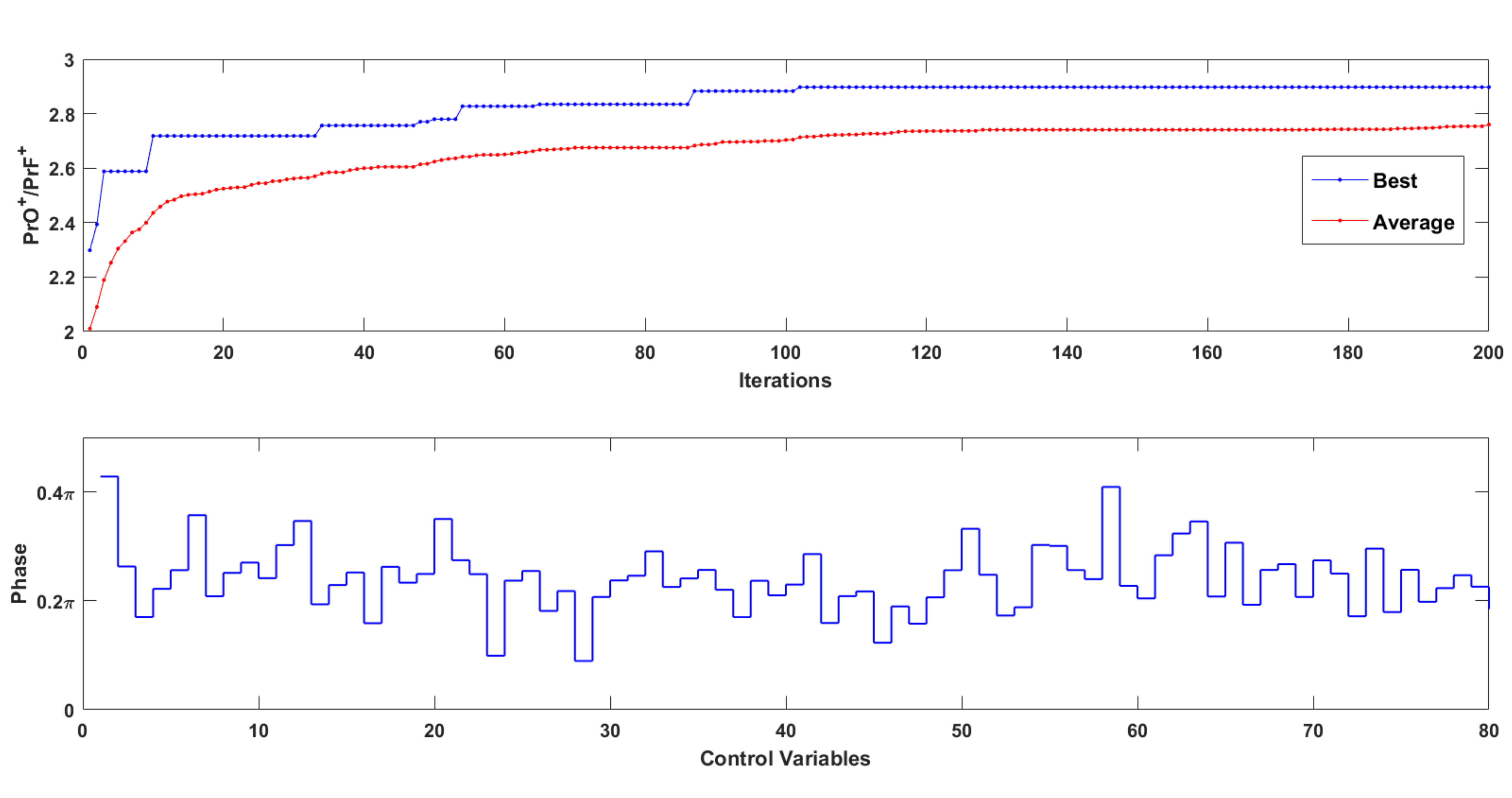}
\caption{Experimental result on the femtosecond laser control system for optimizing the ratio between the products PrO$^{+}$ and PrF$^{+}$ using MS\_DE when the phase is constrained in $[0, \frac{\pi}{2}]$. (a) Ratio PrO$^{+}$/PrF$^{+}$ vs iterations, where `Best' represents the maximum fitness and `Average' represents the average fitness of all individuals during each iteration. (b) Optimized phases of 80 control variables for the optimal result that corresponds to the maximum fitness.}
\label{Experimenthalfpi}
\end{figure*}

\begin{figure*}
\centering
\includegraphics[width=0.80\textwidth]{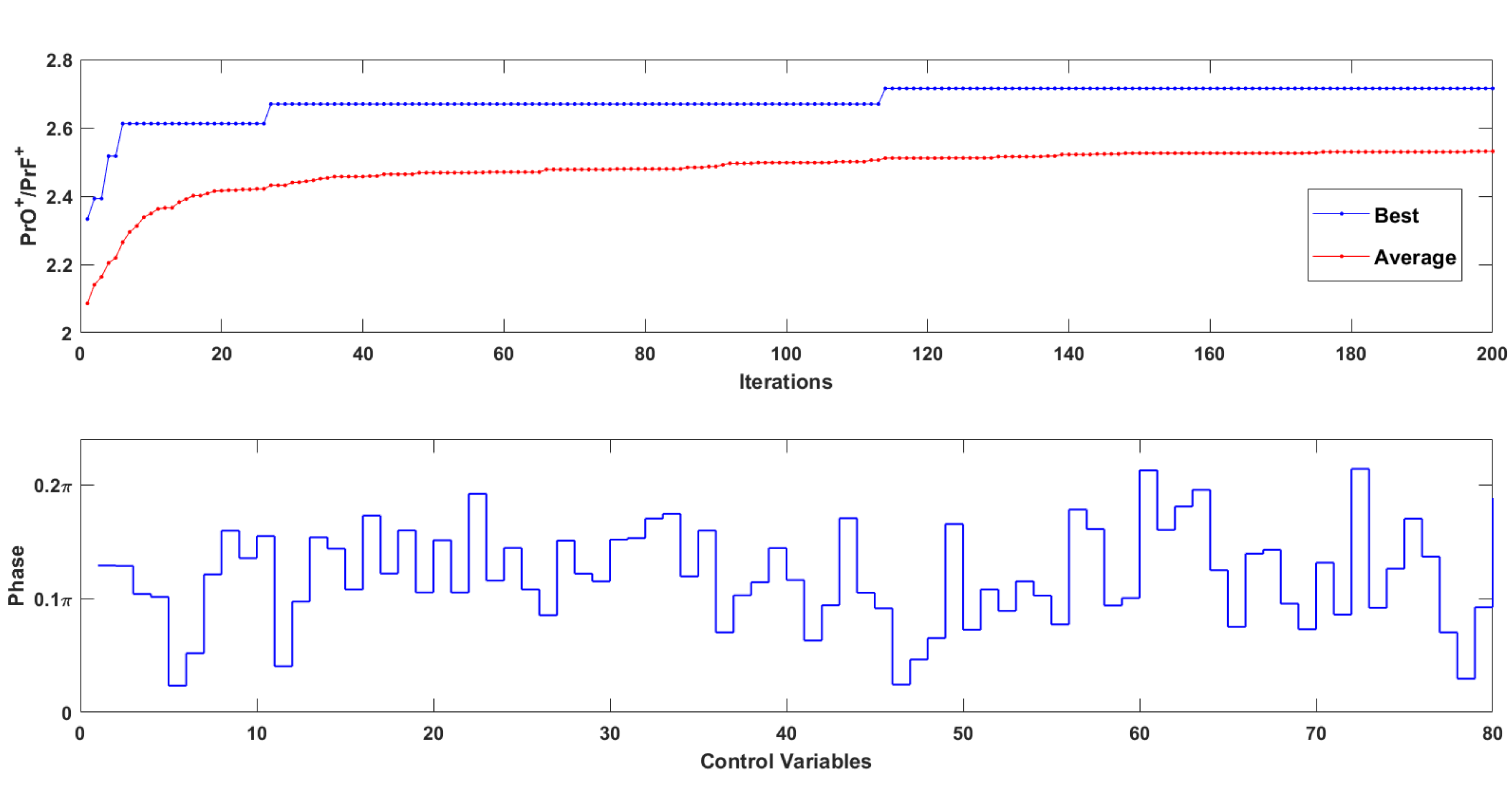}
\caption{Experimental result on the femtosecond laser control system for optimizing the ratio between the products PrO$^{+}$ and PrF$^{+}$ using MS\_DE when the phase is constrained in $[0, \frac{\pi}{4}]$. (a) Ratio PrO$^{+}$/PrF$^{+}$ vs iterations, where `Best' represents the maximum fitness and `Average' represents the average fitness of all individuals during each iteration. (b) Optimized phases of 80 control variables for the optimal result that corresponds to the maximum fitness.}
\label{Experimentquarterpi}
\end{figure*}

\subsection{Experimental setup}
The experiments were implemented on the femtosecond laser control system in the Department of Chemistry at Princeton University. The experimental system consists of three key components: 1) a femtosecond laser system, 2) a pulse shaper, and 3) a time-of-flight mass spectrometry (TOF-MS). In particular, the femtosecond laser system (KMlab, Dragon) consists of a Ti:sapphire oscillator and a amplifier, which produces 1 mJ, 25 fs pulses centered at 790 nm. The laser pulses from the femtosecond laser system are introduced into a pulse shaper that is equipped a programmable dual-mask liquid crystal spatial light modulator. The interaction between the spatial light modulator and the learning algorithm is accomplished by LabVIEW software. The spatial light modulator has 640 pixels with 0.2 nm/pixel resolution and can modulate amplitude and phase independently \cite{Tibbetts JPC}, \cite{Tibbetts PCCP}. Every eight adjacent pixels are bundled together to form an array of 80 ``grouped pixels". Each array of 80 ``grouped pixels" corresponds to a control variable, which can be used to adjust the amplitude and phase values. In these experiments, we consider two constraints: one is on the amplitude values and the other is on the phase values. We fix all the amplitude values at 1 (i.e., fixed energy) for the first constraint, that is, we employ a phase-only control strategy. For the second constraint, we consider the different range of phase values, which may correspond to a situation with magnitude constraint on control inputs. The solid Pr(hfac)$_3$ molecule samples are heated and vaporized into the gas phase in a vacuum chamber with the pressure $1.3 \times 10^{-7}$ Torr. The shaped laser pulses out of the shaper are focused into the vacuum chamber, where photoionization and photofragmentation occurs for the gas-phase Pr(hfac)$_3$ molecules. The fragment ions from these gas-phase Pr(hfac)$_3$ molecules are separated with a set of ion lens and pass through a TOF tube before being collected with a micro-channel plate detector. The mass spectrometry signals are recorded with a fast oscilloscope, which accumulates 3000 laser shots in one second before the average signal is sent to a personal computer for further analysis. A small fraction of the beam ($<5\%$) is separated from the main beam, and focused into a DET25K Thorlab photodiode. The photodiode collects signals arising from two photon absorption for optimizing a given photofragment ratio of Pr(hfac)$_3$ molecules.

\subsection{Fragmentation control}
Before implementing the experiments, we need to optimize the two photon absorption signal to identify the shortest pulse. The process can be used to remove the residual high-order dispersion in the amplifier output. The MS\_DE algorithm is employed to optimize the two photon absorption signal.
Then we consider the fragmentation control of Pr(hfac)$_3$ molecules, where the fitness is defined as the photofragment ratio of PrO$^{+}$/PrF$^{+}$, i.e., $J$=PrO$^{+}$/PrF$^{+}$. We aim to maximize the objective function $J$. The control variables are the phases of femtosecond laser pulses and the MS\_DE algorithm is employed to optimize the phases of 80 control variables. In the learning algorithm, the parameter setting is set as follows: $D = 80$ and $NP = 30$.

In the first experiment, we assume that there are no constraints on the phase values, that is, the phase values may take on arbitrary values between 0 and $2\pi$. An experimentally acceptable termination condition of $1000$ generations (iterations) is used. For $1000$ iterations, it approximately takes twelve hours to run the experiment. For each generation, a total of 30,000 signal measurements are made.
Figure \ref{Experiment2pi} shows the experimental results using the MS\_DE algorithm, where the ratio PrO$^{+}$/PrF$^{+}$ as the fitness function is shown in Fig. \ref{Experiment2pi}(a) and the 80 optimized phases for the final optimal result are presented in Fig. \ref{Experiment2pi}(b). In Fig. \ref{Experiment2pi}(a), `Best' represents the maximum fitness and `Average' represents the average fitness of all individuals during each iteration. With 553 iterations, MS\_DE can find an optimized pulse to make PrO$^{+}$/PrF$^{+}$ achieve 3.067. After 553 iterations, the maximum ratio remains unchanged.

In the second experiment, we assume that the phase values can only vary between 0 and $\pi$. A termination condition of $200$ generations (iterations) has been used to save the experiment time. Figure \ref{Experimentpi} shows the results from the MS\_DE algorithm, where the average ratio PrO$^{+}$/PrF$^{+}$ as the fitness function is shown in Fig. \ref{Experimentpi}(a) and the 80 optimized phases for the final optimal result are presented in Fig. \ref{Experimentpi}(b). MS\_DE can find an optimized pulse to make PrO$^{+}$/PrF$^{+}$ achieve 3.037. Even though the constraint of phase values lying only between 0 and $\pi$, the ratio PrO$^{+}$/PrF$^{+}$ can still reach 99\% of the ratio in the case without phase constraint at 186 iterations.

In two additional experiments, we assume that the phase values can only vary between 0 and $\frac{\pi}{2}$, and between 0 and $\frac{\pi}{4}$, respectively. The termination conditions of $200$ generations (iterations) have been used in the two experiments. The results are shown in Fig. \ref{Experimenthalfpi} and Fig. \ref{Experimentquarterpi}. From \ref{Experimenthalfpi}(a), the MS\_DE algorithm can find an optimized pulse to make PrO$^{+}$/PrF$^{+}$ achieve 2.898 when the phase values are constrained between 0 and $\frac{\pi}{2}$. The ratio PrO$^{+}$/PrF$^{+}$ can achieve 2.715 if the phase values are constrained between 0 and $\frac{\pi}{4}$ as shown in Fig. \ref{Experimentquarterpi}(a). From these results, it is clear that the MS\_DE algorithm can assist in finding good femtosecond laser pulses to optimize the product ratio PrO$^{+}$/PrF$^{+}$ even when different constraints are placed on the amplitude and phase values of the femtosecond laser pulses.



\section{CONCLUSION}\label{sec:Sec7}
We investigated learning control for two classes of ultrafast quantum control problems in the frequency domain where there are constraints on the control fields. When the system model is known, a frequency-domain gradient algorithm can be employed to find optimal control fields. The algorithm has been applied to atomic Rb for selective control of population transfer. When the system model is unknown, a machine learning algorithm can be employed for searching optimal ultrafast pulses. We have experimentally implemented an MS\_DE algorithm in the laboratory to control fragmentation of Pr(hfac)$_3$ molecules with different constraints on ultrafast pulses.



\end{document}